%% file: paper.tex
\documentclass[sigconf]{acmart}
\usepackage{pgfplots}
\usepackage{xcolor}
\usepackage{environ}
\usepackage{pgf-pie}
\usepackage{tabularx}

\pgfplotsset{compat=1.18}
\graphicspath{{./}}

\copyrightyear{2026}
\acmYear{2026}
\setcopyright{cc}
\setcctype{by}
\acmConference[CHI '26]{Proceedings of the 2026 CHI Conference on Human Factors in Computing Systems}{April 13--17, 2026}{Barcelona, Spain}
\acmBooktitle{Proceedings of the 2026 CHI Conference on Human Factors in Computing Systems (CHI '26), April 13--17, 2026, Barcelona, Spain}
\acmPrice{}
\acmDOI{10.1145/3772318.3790682}
\acmISBN{979-8-4007-2278-3/2026/04}

\def\fig#1{Figure~\ref{#1}}
\def\tab#1{Table~\ref{#1}}

\def\sec#1{Section~\ref{#1}}

\NewEnviron{insights}{
  \par\vspace{0.5em}
  \setlength{\fboxsep}{4pt}
  \noindent\fcolorbox{lightgray}{gray!10}{
    \parbox{\dimexpr\linewidth-2\fboxsep-2\fboxrule\relax}{
      \textbf{Key Insights:} \BODY
    }
  }
}

\definecolor{added}{RGB}{0, 104, 150}
\definecolor{removed}{RGB}{196, 49, 25}

\newcommand{\added}[1]{#1}
\newcommand{\removed}[1]{}

\definecolor{likert1}{HTML}{a6611a}
\definecolor{likert2}{HTML}{dfc27d}
\definecolor{likert3}{HTML}{f5f5f5}
\definecolor{likert4}{HTML}{80cdc1}
\definecolor{likert5}{HTML}{018571}

\definecolor{seq1}{HTML}{eff3ff}
\definecolor{seq2}{HTML}{bdd7e7}
\definecolor{seq3}{HTML}{6baed6}
\definecolor{seq4}{HTML}{2171b5}

\begin{document}

\title[Barriers that Programming Instructors Face While Performing
  Emergency Pedagogical Design]{Barriers that Programming Instructors Face
  While Performing Emergency Pedagogical Design to Shape Student-AI Interactions
  with Generative AI Tools}

\author{Sam Lau}
\affiliation{
  \institution{University of California San Diego}
  \city{La Jolla}
  \country{USA}}
\email{lau@ucsd.edu}

\author{Kianoosh Boroojeni}
\affiliation{
  \institution{Florida International University}
  \city{Miami}
  \country{USA}}
\email{kborooje@fiu.edu}

\author{Harry Keeling}
\affiliation{
  \institution{Howard University}
  \city{Washington, DC}
  \country{USA}}
\email{hkeeling@howard.edu}

\author{Jenn Marroquin}
\affiliation{
  \institution{Google}
  \city{Austin}
  \country{USA}}
\email{zap@google.com}

\begin{abstract}
  Generative AI (GenAI) tools are increasingly pervasive, pushing instructors to
  redesign how students use GenAI tools in coursework. We conceptualize this
  work as \emph{emergency pedagogical design}: reactive, indirect efforts by
  instructors to shape student-AI interactions without control over commercial
  interfaces. To understand practices of lead users conducting emergency
  pedagogical design, we conducted interviews ($n$=13) and a survey ($n$=169) of
  computing instructors. These instructors repeatedly encountered five barriers:
  \emph{fragmented buy-in} for revising courses; \emph{policy crosswinds} from
  non-prescriptive institutional guidance; \emph{implementation challenges} as
  instructors attempt interventions; \emph{assessment misfit} as student-AI
  interactions are only partially visible to instructors; and \emph{lack of
    resources}, including time, staffing, and paid tool access. We use these
  findings to present emergency pedagogical design as a distinct design setting
  for HCI and outline recommendations for HCI researchers, academic
  institutions, and organizations to effectively support instructors in adapting
  courses to GenAI.
\end{abstract}

\begin{CCSXML}
  <ccs2012>
  <concept>
  <concept_id>10003120.10003121.10011748</concept_id>
  <concept_desc>Human-centered computing~Empirical studies in HCI</concept_desc>
  <concept_significance>500</concept_significance>
  </concept>
  </ccs2012>
\end{CCSXML}

\ccsdesc[500]{Human-centered computing~Empirical studies in HCI}

\keywords{emergency pedagogical design, student-AI interaction, programming
  instructors, generative AI}

\maketitle

\section{Introduction}
\label{sec:intro}

Instructors around the world are reacting to the availability of highly capable
generative artificial intelligence (GenAI) tools like
ChatGPT~\cite{openaiIntroducing2022}, Claude~\cite{goldmanOpenAI2023}, and
Gemini~\cite{pichaiIntroducing2023}, which have raised questions around the
value of formative assessments like take-home essays and programming assignments
that can easily be completed by these tools~\cite{walshEveryone2025}. Students
already interact with GenAI via freely available GenAI applications, sometimes
even without intending to do so because GenAI outputs are now included by
default in web search results~\cite{reidGenerative2024}, word processing
software~\cite{spataroIntroducing2023}, and code
editors~\cite{hollandAnnouncing2024}. Since unrestricted use of GenAI tools can
hinder student learning~\cite{bastaniGenerative2024a,zhaieffects2024},
instructors are now finding themselves in the unenviable position of trying to
structure student interactions with GenAI tools that weren't originally
designed for instruction.

Drawing an analogy to emergency remote teaching where instructors needed to
abruptly offer courses online in response to the COVID-19
pandemic~\cite{hodgesdifference2020}, we characterize the work that instructors
are now performing as \emph{emergency pedagogical design}. In addition to all of
the previous responsibilities of running a course -- creating course materials,
lecturing, grading, and managing course staff -- instructors now suddenly need
to encourage students to interact with GenAI tools in ways that help rather than
hinder learning. For example, some instructors have banned GenAI usage
entirely~\cite{johnsonChatGPT2023}, while others have found it valuable to
deliberately demonstrate its limitations~\cite{rooseDont2023}. \added{One clear
  similarity between emergency remote teaching and emergency pedagogical design is
  that both arose in response to a disruptive event: in the former, a global
  pandemic, and in the latter, the creation of free-to-use, highly capable GenAI
  tools. One notable difference is that emergency remote teaching was generally
  considered to be a temporary shift in teaching modality until the COVID-19 virus
  was under control, but there is general consensus that GenAI tools will have a
  role in education even in the long-term~\cite{baidoo-anuEducation2023}. Thus, by
  studying instructors' current responses to GenAI tools, we can deepen
  understanding of the practices that may soon be normalized in everyday
  instruction.}

\removed{In this framing, instructors who are designing student-AI interactions share
  some similarities to user experience (UX) designers for software. For example,
  both instructors and UX designers establish goals for a target audience:
  instructors want to maximize student learning, while a designer might want to
  encourage users to engage with their application. To accomplish these goals,
  both groups observe how people interact with software and use this evidence to
  encourage user behavior that is more aligned with their goals. However, there
  are differences as well, as depicted in \fig{fig:overview}. For example, UX
  designers are trained in design thinking and user research methods, while
  instructors primarily focus on pedagogy and their discipline-specific research
  area. But perhaps the most salient difference is that designers often have the
  ability to change the interface itself, while instructors lack control over most
  GenAI interfaces. Instead, instructors manipulate course policies, materials,
  assessments, and classroom culture in order to promote desirable student-AI
  interactions~\cite{kasneciChatGPT2023,digitaleducationcouncilWhat2025}.}

This paper examines undergraduate computing instructors as a critical early case
where GenAI's impact arrived to observe emergency pedagogical design in
practice. In particular, in this setting we consider computing instructors to be
lead users~\cite{vonhippelLead1986} -- users who are trying to design
solutions for their own experienced needs. For example, computing instructors
are attempting to shape student-AI interactions by creating alternative
interfaces that wrap off-the-shelf GenAI tools
(e.g.~\cite{kazemitabaarCodeAid2024,liffitonCodeHelp2023}), making them a
potential source of need-forecasting for instructors in other domains. This
leads to our central research question: \textbf{How do computing instructors
  conduct emergency pedagogical design, and what barriers do they encounter?}

To address this question, we interviewed 13 computing instructors who had made
substantial attempts to influence student-AI interactions by integrating GenAI
into their course assignments, assessments, and infrastructure. To gauge broader
patterns among a more diverse set of computing instructors, \added{including
  instructors at Minority-Serving Institutions~(MSIs) and Historically Black
  Colleges and Universities~(HBCUs)}, we also surveyed 169 instructors about their
perspectives towards integrating GenAI tools into their courses. These
instructors repeatedly encountered five barriers as they engaged in emergency
pedagogical design: \emph{fragmented buy-in} for revising courses; \emph{policy
  crosswinds} from non-prescriptive institutional guidance; \emph{implementation
  challenges} as instructors attempt interventions; \emph{assessment misfit} as
student-AI interactions are only partially visible to instructors; and
\emph{lack of resources}, including time, staffing, and paid tool access.
This study was conducted in mid-2025, about two and a half years after the
initial release of well-known GenAI tools like ChatGPT and GitHub Copilot. Thus,
our work captures a rare moment in time where computing instructors have been
able to make initial attempts to shape student-AI interactions, but approaches
have not yet converged on a set of best practices. To spur the community to use
this unique opportunity to influence how instructors approach student-AI
interaction, we conclude this paper with a set of open research questions for
HCI~(\sec{sec:disc-hci}), recommendations for academic institutions and
funders~(\sec{sec:disc-recs}), \added{and a reflection on lessons learned from
  emergency remote teaching that may also apply in this context~(\sec{sec:disc-ert})}.

In sum, this paper makes the following contributions:

\begin{enumerate}
  \item Conceptualizing emergency pedagogical design through an empirical study
    of computing instructors.
  \item Documenting barriers that instructors encounter as they conduct
    emergency pedagogical design.
  \item Highlighting implications for HCI researchers, academic institutions,
    and funding agencies who wish to support instructors in adapting their
    courses to GenAI.
\end{enumerate}

\section{Related Work}
\label{sec:related}

We organize our review of related work into two areas. We first examine the
broad impact of GenAI on computing education, including instructor perceptions and
reactions. We then review HCI and computing education research that has
developed tools and pedagogical approaches to support productive GenAI use.

\subsection{GenAI Adoption and Policy Responses in Computing Education}
\label{sec:related-adoption}

The rapid development of GenAI models has created both opportunities and
challenges for computing education. In particular, the launch of AI models that
can generate code like GitHub Copilot~\cite{friedmanIntroducing2021} in 2021 and
ChatGPT in 2022~\cite{openaiIntroducing2022} sparked a flurry of research that
found that these tools could solve problems across the undergraduate computing
curricula, for example in introductory programming
courses~\cite{finnie-ansleyRobots2022,reevesEvaluating2023,savelkaCan2023,wermelingerUsing2023,singlaEvaluating2023,savelkaThrilled2023},
data structures courses~\cite{finnie-ansleyMy2023}, programming
competitions~\cite{liCompetitionlevel2022}, and software engineering
courses~\cite{choudhuriHow2024}. Instructors reacted to this in the short-term
by banning the use of GenAI tools in their course policies and reduced the
weight of non-proctored assessments~\cite{lauBan2023}. As these tools became
widely accessible, however, instructors increasingly acknowledged that simply
changing course policies was not enough; the content and learning goals of their
courses also needed to adapt. This change in perspective was motivated by
observations that students used GenAI tools regardless of
policy~\cite{adninExamining2025a,walshEveryone2025,cuScores2023}, that GenAI
coding assistants had already become commonplace in
industry~\cite{mehtaquarter2025, vanianSatya2025}, and that GenAI tools could
provide new opportunities to improve programming
pedagogy~\cite{pratherRobots2023,dennyComputing2024,franklinGenerative2025,raihanLarge2025a}.

Some instructors have begun to integrate GenAI directly into their courses. For
instance, Vadaparty et al. redesigned an introductory programming course to
incorporate GenAI from the start~\cite{vadapartyCS1LLM2024}, and Benario et al.
did the same for a software engineering
course~\cite{gorsonbenarioUnlocking2025}. However, efforts like this seem to be
the exception, not the norm. In a survey of computing instructors in 2024, 75\%
of surveyed faculty believed that their courses' learning objectives should
change in response to GenAI, yet only 35\% reported actually incorporating GenAI
into their courses~\cite{pratherHype2025}. A similar pattern appears outside
computer science: a global survey conducted in early 2025 reported that 88\% of
faculty use AI sparingly or not at all, even though 86\% expect to use AI for
teaching in the future~\cite{digitaleducationcouncilWhat2025}. Reported reasons
for this gap include distrust of
GenAI~\cite{borelaWhat2025,lyuUnderstanding2025,choiInfluence2023,parkPromise2024},
limited understanding of its capabilities~\cite{tanMore2024}, concerns about
effects on learning
outcomes~\cite{zastudilGenerative2023,foroughiDeterminants2024,harveyDont2025},
and concern for potential student
overreliance~\cite{sheardInstructor2024,haqueGenerative2025,elsayaryinvestigation2024,digitaleducationcouncilWhat2025}.

In contrast to this line of prior work that focused on instructor intentions or
perceptions, we studied what instructors actually did in practice to
change their course materials and report experienced barriers in addition to
perceived barriers.

\subsection{Systems that Support Learning to Program with GenAI}
\label{sec:related-systems}

Since GenAI use for programming has become more common in the past few years,
research on how novices learn programming with GenAI is in its early stages with
both positive and negative results. For instance, Kazemitabaar et al. found that
using a GenAI coding assistant can improve novices' ability to write code even
on later tasks without access to GenAI~\cite{kazemitabaarStudying2023}. However,
GenAI tools impose additional metacognitive requirements on
users~\cite{tankelevitchMetacognitive2024}, which can reinforce learning
challenges for struggling students in the context of
programming~\cite{pratherWidening2024}.

HCI research has proposed system designs to mitigate the potential harms of
GenAI coding tools while maximizing their benefits for programming learners. For
example, GenAI coding tools are often designed to quickly produce working
code~\cite{jimenezSWEbench2024}, not to help learners understand how programs
work. In response, researchers have proposed GenAI tools with guardrails that
offer conceptual help rather than writing
code~\cite{kazemitabaarCodeAid2024,liffitonCodeHelp2023}, and that ask
students to ``teach'' a chatbot~\cite{rogersPlaying2025a}.
To improve user understanding of code, Ivie adds lightweight explanations of
generated code~\cite{yanIvie2024}, Leap surfaces live runtime
values~\cite{ferdowsiValidating2024a}, and WaitGPT visualizes the flow of data
through a program~\cite{xieWaitGPT2024}.

Novices can struggle with writing effective
prompts~\cite{zamfirescu-pereiraWhy2023} and often omit important detail in
their prompts when trying to generate code~\cite{lucchettiSubstance2024}. In
contrast, systems that guide users to decompose problems can increase perceived
control and ease of use~\cite{kazemitabaarImproving2024}. Along this line of
work, DBox guides learners to decompose their algorithms before writing
code\cite{maDBox2025a}, CoLadder provides an interface to manipulate a tree of
prompts and generated code~\cite{yenCoLadder2024}, iGPT iteratively asks
followup questions to learners to produce more complete program generation
plans~\cite{yehBridging2025}, and Prompt Problems automatically grade learner
prompts by checking generated code against test cases~\cite{dennyPrompt2024}.
Some systems even remove the need for prompting altogether by offering proactive
suggestions~\cite{chenNeed2025,khuranaIt2025} or by offering a simplified
interface to generate hints~\cite{roestNextStep2024}.

Together, this prior work advances systems that aim to improve student-AI
interactions. We complement these system-centric contributions by framing
instructors' adoption efforts as the process of emergency pedagogical design,
and by identifying instructor needs that past work has not yet addressed in
order to motivate further HCI research.

\section{Methodology}
\label{sec:methods}

To understand the characteristics of emergency pedagogical design, we conducted
semi-structured interviews with 13 full-time instructors teaching
undergraduate-level computing courses. We were specifically interested in
understanding prior experiences of designing student-AI interactions rather than
merely perceptions of possible interventions. To accomplish this, we selected
interviewees who self-reported that they had made changes to their course
materials because of student-AI usage, and had also taught these updated
materials to students already. We filtered out instructors who had not yet made
changes to their course or had only made changes to their course policies, since
we felt this group of instructors had already been sufficiently studied in prior
work (e.g.~\cite{lauBan2023}). Interviewees were recruited via professional
networks, email lists for computing instructors, and personal contacts, using
snowball and purposive sampling~\cite{palinkasPurposeful2015} that was iterative
until saturation. The interviews were conducted in May and June 2025. All
interviews were held in English and participants were not financially
compensated for participating in the study.

\subsection{Interview Protocol}
\label{sec:protocol}

Each interview was conducted over Zoom by one researcher, lasted between 45-60
minutes, and was audio-recorded upon obtaining consent. Our interview protocol
began with three background questions:

\begin{enumerate}
  \item What programming courses have you most recently taught?
  \item From your perspective, have GenAI tools like GitHub Copilot and ChatGPT
    had an impact on this course? If so, what kinds of impact have you observed?
  \item Does your course or department have a formal policy regarding student
    use of GenAI tools?
\end{enumerate}

The purpose of these background questions was to establish the context of our
interviewees' instruction as of mid-2025, approximately 2.5 years after the
initial release of popular GenAI tools like ChatGPT and GitHub Copilot. These
questions led into the primary open-ended scenario which took up the majority of
our interviews:

\begin{quote}
  Please open and screen-share course materials from a memorable assignment,
  assessment, or moment in your course where you incorporated GenAI tools. Could
  you explain how you wanted students to interact with GenAI in this part of the
  course?
\end{quote}

As part of our protocol, we prepared followup questions to understand instructor
motivation for making this change to their course materials, their experiences
implementing and evaluating this change, and student reactions. We also asked
interviewees followup questions to help elicit more thorough responses.

\subsection{Rationale for Our Interview Protocol}
\label{sec:rationale}

Our protocol design was based on several theoretical considerations. Since we
were specifically interested in what instructors actually did (in contrast to
what they wanted to do), we drew from critical incident
technique~\cite{flanagancritical1954}, a qualitative method for collecting and
analyzing events that had substantial impact on outcomes. In our context, this
means that we spent the majority of the interview asking instructors to reflect
on the most memorable interventions that they performed in their courses, with
the expectation that talking about vivid experiences would encourage
interviewees to share more relevant details. Note that we did not define
``memorable'' as ``positive''; we were equally interested in unsuccessful
experiences attempting to design student-AI interactions and made this clear to
our interviewees as part of our protocol.

We also grounded our conversations around concrete pieces of course materials,
such as take-home assignments or exam questions. This part of the protocol was
inspired by the cognitive walkthrough~\cite{mahatodyState2010}
methodology in HCI, which recommends discussing a shared artifact to help elicit
greater quality and quantity of observations. To reduce the risk of interviewees
fixating on minute details of their course materials, we also asked interviewees
to make higher-level reflections after each line of questioning.

Lastly, when designing individual interview questions, we applied best practices
to reduce bias, such as using neutral wording and avoiding
jargon~\cite{adamsConducting2015}. In our protocol, we also did not mention
specific AI tools to avoid priming or anchoring biases, but if participants
brought up a specific tool of their own accord, we asked followup questions
about that tool.

\subsection{Interview Participant Backgrounds}
\label{sec:backgrounds}

\input{tab-participants.tex}

Our interviewee backgrounds are summarized in \tab{tab:participants}. We
recruited 13 participants (5 non-male). Most of our participants taught in the
United States (10/13), while others taught in Canada (1/13), New Zealand (1/13),
and the United Kingdom (1/13). \added{All participants held PhD degrees and were
  either tenure-line faculty (8/13) or nontenure-line faculty (5/13). With one
  exception that taught in an engineering department, all interviewees taught in
  computer science departments.} Interviewees
taught undergraduate courses \added{at both introductory and upper levels},
including introductory programming, data structures, algorithms, software
engineering, and cybersecurity. There was considerable range in the number of
years of full-time teaching experience (5--42 years, $\mu$=17 years).
Participants primarily taught at public PhD-granting universities (10/13), while
some taught at private PhD-granting universities (2/13) and undergraduate-only
institutions (1/13).

\subsection{Survey}

Since our interviewees were selected specifically because they had already
conducted interventions to improve student-AI interactions, we recognized that
our interviews would likely emphasize perspectives of instructors who wanted to
keep including GenAI usage as part of their courses. In order to reduce the
impact of this limitation, in parallel with our interviews we conducted a survey
to gather perspectives of computing faculty more broadly, including faculty that
had not changed their course materials in response to GenAI usage. The survey
was designed to be completed within 10 minutes and asked participants questions
about their personal views, their perceptions of colleagues, and basic teaching
demographics such as average number of courses taught per year and number of
students taught per year.

The survey was distributed via email lists for computing faculty, discussion
forums, and personal outreach. To capture viewpoints from more diverse faculty,
we specifically contacted email lists for faculty at Minority-Serving
Institutions~(MSIs)~\cite{liCharacteristics2007} and Historically Black Colleges
and Universities~(HBCUs)~\cite{allenHistorically2007}.
A total of 169 faculty responded to the survey. More than half of these faculty
taught at MSIs~(51\%), and a sizable proportion taught at HBCUs~(17\%). Most
faculty either taught 1-2 courses per academic term~(37\%) or 3-4 courses per
academic term~(41\%), although a small minority taught more than 5 courses per
term~(17\%). \added{Because the survey relied on convenience and snowball
  sampling across lists with unknown membership sizes, and because respondents
  could skip any item, we could not calculate a response rate; instead, we treat
  the results as descriptive of respondents rather than as estimates of wider
  faculty attitudes. }

\subsection{Data Analysis}

During each interview, the researcher who conducted the interview took
timestamped notes. Then, a second researcher independently listened to each
audio recording and took their own set of notes. Both researchers met weekly to
discuss their notes together. After the interviews were completed, we
iteratively came up with themes using an inductive
approach~\cite{corbinBasics2014}. \added{Both researchers independently
  annotated interview excerpts at the level of short utterances or question-answer
  turns, depending on where a shift in meaning occurred. We then compared
  annotations in weekly meetings. Coding proceeded as a negotiated process rather
  than through formal inter-rater reliability statistics: disagreements were
  discussed case by case, and codes were revised until both researchers reached
  consensus. This process unfolded over five weekly iterations, during which we
  merged, split, and renamed codes until the codebook stabilized. After
  convergence, we re-reviewed earlier excerpts to ensure consistent application.
  This iterative process led us to center each theme on barriers instructors
  encountered while attempting to alter student-AI interactions.
}
\removed{we annotated interview quotes with potential
  themes, then merged and split themes based on our weekly discussions until we
  converged on our final set. During this process, we decided to center each of
  our themes on a barrier that instructors face while attempting to alter
  student-AI interactions because of the prevalence of repeated barriers across
  our interviewees.}

To analyze the survey data, two researchers computed counts and proportions for
Likert-scale questions and applied an inductive approach to generate themes for
open-ended responses. \added{For open-ended items, both researchers
  independently labeled sentences within each response. We compared labels in
  weekly meetings and resolved disagreements through discussion until agreement
  was reached. Coding stabilized after three iterations, after which we
  re-reviewed earlier responses for consistency and grouped the agreed-upon codes
  into the themes reported in the results.} Although we allowed survey
participants to leave any question blank if they chose, we found that all of our
Likert-scale questions had response rates over 90\% (between 152-161 responses
out of 169 submissions). As such, when reporting Likert-scale question results,
we omit the number of responses for ease of reading.

\subsection{Study Scope and Limitations}

We purposively recruited interviewees who had already made changes to student-AI
interactions beyond policy-only edits. This focus may bias the interviews toward
instructors with more time, staffing, or institutional support and
under-represent instructors who chose not to engage or were unable to do so. By
asking for memorable incidents and successful or attempted interventions, the
data may overweight vivid cases and underweight routine or abandoned efforts.
Accounts are retrospective self-reports and thus subject to recall and
social-desirability biases, so we treat quotations as situated evidence rather
than measurements. Despite efforts to recruit broadly, interviews were conducted
in English and were predominantly with U.S.-based faculty at research-intensive,
PhD-granting institutions; transferability to community colleges,
teaching-focused institutions, and non-U.S. contexts (including non-English
instruction) may be limited.

The survey used convenience and snowball distribution via email lists, forums,
and personal outreach rather than probability sampling. We did not weight
responses and allowed missing items, so we cannot estimate population parameters
or make inferential claims beyond respondents. \added{This approach likely
  introduced self-selection biases: instructors with strong views about GenAI,
  whether positive or negative, may have been more motivated to respond.} Although
we sought to include a more diverse range of responses from MSIs and HBCUs, the
sample is nonetheless not nationally representative. Because of these
limitations, in this paper we treat survey results as descriptive of our
respondents and use percentages mainly to compare or contrast with interview
patterns rather than as prevalence estimates.

\added{Because all interviewees taught programming-focused courses and the
  survey focused on computing faculty, our findings are more directly relevant
  within computing education and may not generalize to other domains. We
  attempted to frame our findings so that they have the potential to appear in
  other disciplines, but we acknowledge that instructors in writing-intensive,
  humanities, or social science contexts may face distinct challenges that our
  data do not capture. As such, transferability of these findings beyond
  computing requires caution, and we view this work as an early case in one
  discipline where GenAI impacted teaching practices rapidly.}

\section{Results}
\label{sec:results}

\input{tab-results-overview.tex}

In this section we present a definition and properties of emergency pedagogical
design as instructors adapted their courses to account for student GenAI
usage~(\sec{sec:definition}). During this design work, instructors faced five
barriers that repeatedly arose during our analysis: fragmented
buy-in~(\sec{sec:personal}); policy crosswinds~(\sec{sec:administrative});
implementation challenges~(\sec{sec:student-ai}); assessment
misfit~(\sec{sec:assessment}); and lack of resources~(\sec{sec:resources}).

\added{Both the properties of emergency pedagogical design and the five barriers
  reported in this section were derived from the same inductive thematic analysis
  described in Section~\ref{sec:methods}. During coding, we noted two kinds of
  recurring patterns: cross-cutting characteristics of the design setting that we
  refer to as properties of emergency pedagogical design, and recurring challenges
  that we grouped into five barriers. The properties characterize the conditions
  under which instructors worked, while the barriers describe the constraints that
  shaped their attempts to change student-AI interactions.}

\subsection{Overview of Emergency Pedagogical Design}
\label{sec:definition}

In the context of our instructors reacting to student usage of GenAI tools, we
define \emph{emergency pedagogical design} as the work that instructors do
to adjust their courses in order to improve student interactions with GenAI.
\added{In synthesizing interview accounts, we observed several cross-cutting patterns
  that described the broader conditions under which instructors tried to adjust
  their courses. We refer to these as properties of emergency pedagogical design
  because they characterize the design setting that emerged from our inductive
  analysis rather than specific challenges or solutions.} We observed the following
properties:

\begin{itemize}
  \item \textbf{Reactive rather than proactive:} Instructors did not anticipate
    the capabilities or extent to which students would use GenAI tools without
    instructor intervention. Since the instructors in our study had been
    teaching existing courses, instructors generally had to retrofit course
    materials that were created before GenAI tools were widely used rather than
    write new materials from scratch that were designed to account for GenAI
    usage.
  \item \textbf{Indirect rather than direct:} Interaction designers typically
    have the ability to make direct changes to the interface they are designing.
    In contrast, instructors lack the ability to directly update interfaces of
    GenAI tools, since these tools are maintained by corporations rather than
    academic institutions. In addition, GenAI output is increasingly embedded
    within everyday interfaces (e.g., web search~\cite{reidGenerative2024} and
    code editors~\cite{hollandAnnouncing2024}), making it challenging for
    instructors to anticipate all the ways that students can trigger GenAI
    usage, whether intentionally or inadvertently. Without the ability to change
    GenAI interfaces directly, instructors indirectly influence student
    interactions with GenAI through learning materials, course policies,
    assessment strategies, and course infrastructure.
  \item \textbf{Ambient evidence rather than orchestrated evaluations:} Rather
    than in-depth user studies, instrumentation, or A/B tests, instructors
    inferred effects from informal signals such as office-hour conversations,
    anecdotes from course staff, help-queue and autograder patterns, forum
    threads, and shifts in attendance.
  \item \textbf{Timely iteration rather than deferred certainty:} Instructors
    felt that they couldn't wait for cumulative research evidence or stable
    tools to emerge; they instead wanted to act quickly, making minimally viable
    policy and curricular changes and then iterating based on local signals.
\end{itemize}

Taken together, these properties portray emergency pedagogical design as
reactive, indirect, and iteration-driven work conducted under uncertainty and
with limited visibility into student behavior. In the following subsections, we
elaborate on five recurring barriers that constrained how instructors carried
out emergency pedagogical design within this setting, summarized in
\tab{tab:results-overview}.

\subsection{Fragmented Buy-In}
\label{sec:personal}
\begin{figure}
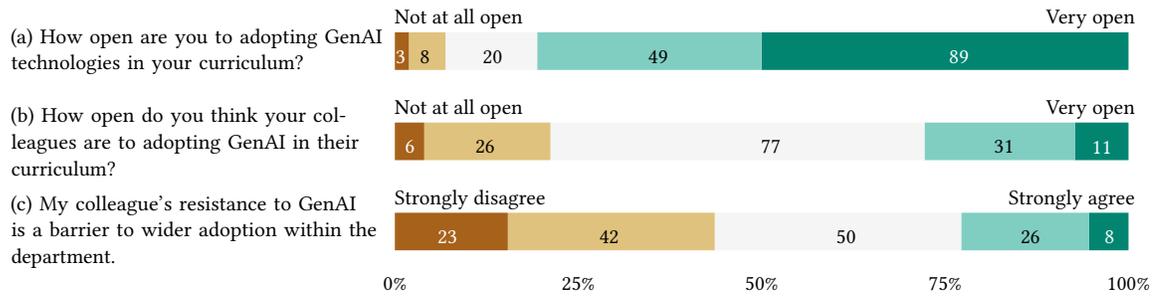

  \centering
  \include{fragmentedbuyin}
  
  \caption{Although a strong majority of survey respondents considered
    themselves open to adopting GenAI technologies (a), only a minority felt
    their colleagues were equally open (b). However, only a few instructors perceived
    colleague resistance to GenAI as a barrier (c), suggesting that ambivalence was
    more common than direct opposition. Instructors responded on a 5-point
    scale. The number of responses for each option is displayed within each
    bar.}
  \Description{Stacked bar plots showing responses to three survey questions:
    (a) How open are you to adopting GenAI technologies in your curriculum?; (b)
    How open do you think your colleagues are to adopting GenAI in their
    curriculum?; and (c) My colleague's resistance to GenAI is a barrier to wider
    adoption within the department.}
  \label{fig:fragmentedbuyin}
\end{figure}

We use \emph{buy-in} to refer to two levels of commitment: an instructor's own
motivation to revise course materials to shape student-AI interactions, and
departmental support from colleagues that enables and rewards those revisions.
\added{We make this distinction because our survey results indicated a clear gap
  between instructors' own openness to adopting GenAI and their perceptions of
  departmental support. While 81\% of respondents rated themselves as Open or Very
  Open to adopting GenAI technologies, only 28\% said the same of their colleagues
  (Fig.~\ref{fig:fragmentedbuyin}a,b). Although 23\% Agreed or Strongly Agreed
  that colleague resistance was a barrier, most respondents selected neutral
  options, suggesting ambivalence rather than direct opposition
  (Fig.~\ref{fig:fragmentedbuyin}c). These patterns indicate that instructors
  often felt individually willing to revise courses but did not perceive a shared
  departmental commitment to this work.}

\removed{Emergency pedagogical design competed with routine teaching work.
  Instructors were already committed to preparing and updating assignments,
  delivering lecture, meeting with staff, answering email, writing exams, and
  grading. Managing unprompted student use of generative AI arrived on top of
  these duties, not in place of them. As a result, designing student-AI
  interactions was rarely their top priority, even when instructors believed it
  mattered for learning.}

Since we selected for interviewees that had already made concrete changes to
their course materials in response to student GenAI usage, \added{a limitation
  of our interviews is that they only surface perspectives of instructors who
  had strong personal buy-in for making course changes.} Even so, several
interviewees described an initial period of reluctance or skepticism~(P03-P06,
P09-P10, P12). For example, one instructor mentioned that ``I'd been trying to
ignore [GenAI tools] for as long as I possibly could'' by using proctored
quizzes to deter students from using GenAI on assignments~(P06). Despite having
course policies forbidding GenAI, this instructor eventually realized that
students used GenAI to complete assignments anyway, and then did poorly on
proctored assessments -- a pattern echoed by other interviewees~(P04, P08, P12,
P13). On reflection, P06 shared that ``punishing them [via a proctored
    assessment] after the fact is not particularly useful or effective. So how can I
get them to stop doing things they shouldn't be doing?'' Other instructors cited
a desire to curb potential ``bad habits'' when students used GenAI without
instructor guidance~(P07, P08).

Beyond personal buy-in, departmental buy-in was also fragmented. Interviewees
reported that they were in a small minority within their departments who were
meaningfully integrating AI into coursework (P01, P05, P07, P08, P11), \added{a
  pattern that aligned with our survey data}. Some even faced opposition from
other faculty in their department about trying to include AI interactions in
coursework~(P05, P10-P12). The mixed reactions by other faculty in their
departments reduced informal support, slowed coordination across multi-section
courses, and made it harder to share materials or run aligned evaluations. This
isolation also constrained scope: all of our interviewees piloted
course-specific changes, but were not yet able to attain broader alignment with
other courses via shared policies, common assignment patterns, or cross-course
scaffolds.

\removed{This fragmentation was echoed outside of our interviewees -- although 81\% of
  our survey respondents said that they were either Open or Very Open to adopting
  AI tools in their teaching, only 28\% said the same about their colleagues
  ~(\fig{fig:fragmentedbuyin}a,b). Instructors reported colleague ambivalence rather than
  direct opposition, since only 23\% Agreed or Strongly Agreed with the statement:
  ``My colleagues' resistance to using GenAI tools in their courses is
  a barrier to wider adoption at my institution.''~(\fig{fig:fragmentedbuyin}c).}

\begin{insights}
  Together, limited personal bandwidth and weak departmental support narrowed
  what was feasible. Instructors tended to adopt small, local changes they could
  own, rather than larger designs that required coordination, new tooling, or
  sustained evaluation.
\end{insights}

\subsection{Policy Crosswinds}
\label{sec:administrative}

\begin{figure}
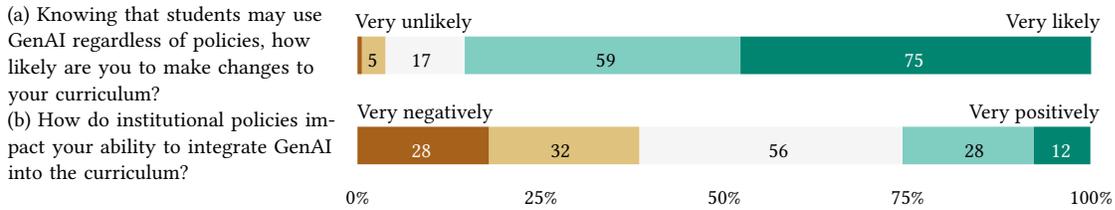

  \centering
  \include{policycrosswind}
  \caption{Surveyed instructors felt motivation to update their curriculum in
    response to student usage of GenAI (a). Institutional constraints had mixed
    perceived impact (b), \added{yet unequal access to paid GenAI tools was a
      widely held concern (c).} Instructors responded on a 5-point scale. The
    number of responses for each option is displayed within each bar. }
  \Description{Stacked bar plots showing responses to three survey questions:
    (a) Knowing that students may use GenAI regardless of policies, how likely
    are you to make changes to your curriculum?, (b) How do institutional
    policies impact your ability to integrate GenAI into the curriculum?, (c) To
    what extent do you agree or disagree with the following statement: "Unequal
    access to paid generative AI tools among students could create or exacerbate
    disparities in learning outcomes and workforce preparedness."}
  \label{fig:policycrosswind}
\end{figure}

\added{Since a growing number of institutions have started to issue guidelines
  about GenAI usage in coursework~\cite{mcdonaldGenerative2025}, our survey
  asked computing instructors about the effects of these top-down policies on
  their adoption of GenAI.} Knowing that students may use GenAI tools regardless
of policies forbidding them, 85\% of respondents said they were Likely or Very
Likely to change their curriculum~(\fig{fig:policycrosswind}a). By contrast,
when asked how institutional constraints or policies affected their ability to
integrate GenAI, only 26\% reported a Positive or Very Positive
impact~(\fig{fig:policycrosswind}b). \added{Respondents also expressed
  concerns about the uneven enforceability of institution-level policies, noting
  in open-ended responses that policies rarely specified how GenAI should be
  incorporated into assignments, how to interpret allowed versus disallowed
  uses, or how students' access to paid tools should be handled.}

Likewise, nearly all interviewees reported that neither their department nor
institution provided prescriptive policies governing GenAI use in teaching as a
result of fragmented buy-in~(P01-P08, P10-P13). In the absence of top-down
policies, instructors had to decide their own policies on a per-course basis. As
a result, practices diverged, sometimes dramatically. As one instructor
summarized: ``From a student perspective, it's the wild west. Some courses allow
GenAI usage, some don't.''~(P07). Instructors pointed out that this caused
``lots of confusion''~(P01) for students who had to navigate different policies
for every course that they took. Although several instructors mentioned
departmental meetings to discuss GenAI policies, discussions had not yet
converged shared guidelines or consistent approaches~(P01, P02, P04, P05, P07,
P10, P12). These inconsistencies also resulted in equity implications that
current policies often did not address. For example, none of our interviewees
reported course policies that distinguished between access to paid versus unpaid
tools or between standalone chatbots and GenAI embedded in everyday software
(e.g., code editor assistants or web search). \added{The difference between paid
  and unpaid tools in particular was salient among survey respondents: 78\% of
  responses Agreed or Strongly Agreed with the statement, ``Unequal access to paid
  generative AI tools among students could create or exacerbate disparities in
  learning outcomes and workforce preparedness''~(\fig{fig:policycrosswind}c).}

As one notable counterexample, P09 described a university ``two-lane'' policy
that distinguished proctored from unproctored assessments. For proctored
assessments, instructors could set GenAI rules; for unproctored assessments like
take-home programming exercises, ``students can use whatever tools they choose,
and instructors cannot prohibit GenAI use''~(P09). This instructor voiced
support for this approach as a pragmatic way to align policy with what could be
reasonably enforced outside proctored settings.

\removed{
  The survey echoed these crosswinds. Knowing that students may use GenAI
  regardless of policies, 85\% of respondents said they were Likely or Very Likely
  to change their curriculum~(\fig{fig:policycrosswind}a). By contrast, when asked how
  institutional constraints or policies affected their ability to integrate GenAI,
  only 26\% reported a Positive or Very Positive
  impact~(\fig{fig:policycrosswind}b).
}

\begin{insights}
  Inconsistent or absent guidance produces divergent course policies. Students
  must track different rules across classes, and course policies rarely account
  for paid versus unpaid tools or GenAI embedded in common software, raising
  equity concerns.
\end{insights}

\subsection{Implementation Challenges}
\label{sec:student-ai}

\begin{figure}
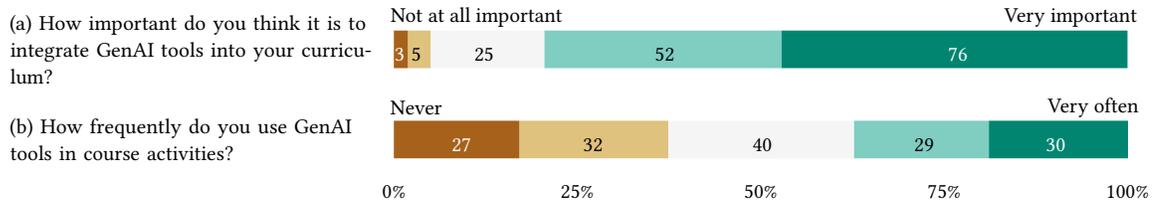

  \centering
  \include{implementationchallenges}
  \caption{Surveyed instructors felt it was important to integrate GenAI tools
    into their curriculum (a), but fewer instructors reported actually doing so
    (b). Instructors responded on a 5-point scale. The number of responses for
    each option is displayed within each bar.}
  \Description{Stacked bar plots showing responses to two survey questions:
    (a) How important do you think it is to integrate GenAI tools into your
    curriculum?, (b) How frequently do you use GenAI tools in course activities? }
  \label{fig:implementationchallenges}
\end{figure}

Instructors wanted to shape how students used GenAI, but their leverage was
indirect. They could adjust policies, task design, workflows, and course
infrastructure, but not the commercial tools students actually used. We observed
a spectrum of redesigns. At the lightest end, some instructors kept existing
bans but permitted GenAI in specific programming labs or assignments~(P03, P06,
P13). At the most extensive end, others told students to use GenAI across most
assignments, integrating it into routine student practice~(P01, P02, P05, P08,
P11). Some instructors even built custom interfaces for students that used GenAI
tools as part of their implementation~(P04, P09).

One way to frame instructor actions is to describe them as interventions that
are intended to encourage certain student behaviors with GenAI tools. We
highlight selected examples to demonstrate the diversity of instructor
approaches:

\begin{itemize}
  \item \emph{Prompting as decomposition (P01).} {Intervention:} verbally
    encouraged students to use GenAI to generate small, concrete inputs for
    algorithms. {Intended behavior:} prompt at the level of subproblems
    rather than paste full problem statements.

  \item \emph{Tasks that resist one-shot answers (P03).} {Intervention:}
    designed a lab with a malformed CSV and multiple valid approaches.
      {Intended behavior:} recognize that GenAI outputs can be unreliable and
    require validation.

  \item \emph{Guardrailed conceptual help (P04).} {Intervention:}
    deployed a custom chatbot that guided reasoning but avoided giving code.
      {Intended behavior:} have students write code themselves while using
    GenAI for conceptual support.

  \item \emph{Underspecified requirements and judgment (P06).} {Intervention:}
    included a problem where a critter's score must converge but ``it may wiggle
    around a bit.'' {Intended behavior:} practice converting an ambiguous
    specification to a precise prompt.

  \item \emph{Refactoring with intent (P07).} {Intervention:} showed that
    GenAI refactors better when students name target designs (e.g., ``replace the
    if statements with an abstract interface''). {Intended behavior:} motivate
    learning of foundational concepts to work productively with GenAI.

  \item \emph{Plan before you generate (P05, P08).} {Intervention:} required a
    design document before asking GenAI to implement. \emph{Intended behavior:}
    keep planning and decomposition as student responsibility rather than
    outsourcing the whole project.

  \item \emph{Explanations as first-class work (P09).} {Intervention:} built a
    pipeline that autograded code produced from student explanations of code via
    an AI tool. {Intended behavior:} position GenAI as a tool for learning
    activities (explaining, testing), not just writing code.

  \item \emph{GenAI as a component, not a destination (P12).}
    {Intervention:} had students call GenAI as an API within a program.
      {Intended behavior:} view GenAI as one element in a larger system rather
    than a standalone chatbot.
\end{itemize}

We interpret the wide range of instructor responses to GenAI usage as evidence
of the perceived instructor need to intervene even without clear guidelines from
policy or research on effective student-AI interactions. Interviewees found
navigating this uncertainty challenging. As one instructor said, ``we have to
figure it out on our own''~(P08).

Survey results suggest that our interviewees are not enthusiastic outliers about
teaching GenAI usage in their coursework: 80\% rated integrating GenAI into
computing curriculum as Important or Very
Important~(\fig{fig:implementationchallenges}a). However, fewer survey respondents
had already begun this process, as only 37\% used GenAI tools during course
activities Often or Very Often~(\fig{fig:implementationchallenges}b).

\begin{insights}
  Lacking direct control over commercial tools and operating under sparse
  guidance, instructors needed to work through policies, task design, workflow
  requirements, and custom wrappers to shape \emph{how} students used GenAI
  rather than \emph{whether} they used it.
\end{insights}

\subsection{Assessment Misfit}
\label{sec:assessment}

We use ``assessment'' in two senses that instructors treated as related but
distinct: (1) behavioral alignment: whether students used GenAI in the ways
instructors intended; and (2) learning without scaffolds: whether students could
explain, trace, and write code unaided. In most reported cases, interviewees
used proctored assessments as a way to check whether students were interacting
with GenAI tools in intended ways. However, understanding student behavior prior
to a summative assessment was challenging because instructors only had partial
visibility into day-to-day tool use.

\paragraph{Challenges of understanding student-AI interactions}
Instructors observed an assignment--exam gap: students performed well on
auto-graded or take-home work yet struggled on proctored tasks~(P04, P06, P08,
P10, P12, P13). This motivated them to examine student-AI interactions more
closely and attempt to make changes. P04 reported a memorable example during the
quarter where he first introduced his custom chatbot. On a mid-quarter skill
demo requiring students to write a short Python function from scratch, ``One
third of my students got a 0. This was a class of 450 students, so 150 students
got a 0, which was very concerning to me.'' P04 had originally planned to
introduce students to GitHub Copilot and allow them to use Copilot for the
remainder of the course, but decided to completely backtrack and not introduce
Copilot at all because he was concerned about student knowledge of programming
basics.

P08 described a related pattern: a confidence--competence gap. Students who had
succeeded on take-home work ``were very insistent that they understood the
material, yet when I asked them followup questions about their code, they could
not respond.'' P08 suspected these students were using copy-paste workflows and
limited practice with writing or tracing code from scratch but could not verify
these behaviors across the full class. P08's example illustrates a shared
challenge with assessing student-AI interactions. Because student-AI
interactions occurred in commercial tools outside course infrastructure by
default, instructors lacked reliable telemetry: they could not see which tools
were used, how prompting evolved, or whether course guidelines were bypassed. As
a result, instructors inferred effects from ambient signals -- office hours,
staff anecdotes, autograder patterns, and proctored performance -- rather than
from planned, instrumented evaluations.

Even in cases where this telemetry was possible in theory, understanding student
behavior in practice was difficult. Since students used his custom chatbot, P04
had access to logs of student interactions. However, he noted that analyzing
hundreds of conversation logs was impractical during the term since it would
take too much time away from routine teaching work. P07 shared an experience
building a similar custom chatbot, only to find that students found the chatbot
less helpful than commercial tools and chose to use commercial tools instead.

\paragraph{Challenges of assessing student knowledge}
Other instructors spent considerable effort redesigning their course's
assignments and grading criteria. For example, P02 shifted most assignment
credit to brief, weekly ``stand-up'' meetings where students demonstrated and
discussed their own code. Submitted code counted for only 15--20\% of the grade;
these weekly meetings carried the bulk of credit instead. Stand-up meetings
targeted three behaviors: communicating the program; explaining how data flows
through the program; and predicting outcomes under small changes. P06 adopted
similar standup meetings, noting that these meetings quickly surfaced students
who were struggling and deterred integrity problems. A shared challenge was
ensuring that grading was consistent across all students. In this case, both P02
and P06 felt the need to create detailed, step-by-step instructions for their
course staff on how to grade every assignment that used stand-up meetings.

In an algorithms course, P01 awarded full credit for ``good-faith effort'' on
weekly problem sets and scaled the amount of feedback with the amount of visible
work. This grading emphasized engagement and positioned feedback as the main
incentive, while summative written exams still assessed individual reasoning.

P13 reoriented some assignments so that a small portion involved obtaining GenAI
output, with most credit tied to students' critiques of that output (e.g.,
identifying inconsistencies and justifying revisions), shifting assessed value
from production to evaluation.

\paragraph{Tensions and open questions.} These redesigns raised four unresolved
issues. First, \emph{validity}: do orals, critiques, and explanation items
measure the target constructs (e.g., program comprehension)? Second,
\emph{reliability}: instructors noted questions about TA calibration for orals
and alignment between AI-graded and human-graded judgments but did not report
formal checks. Third, \emph{feasibility}: weekly standup meetings and detailed
assignment feedback scaled only with substantial staffing effort. Fourth,
\emph{equity}: when students differ in access to paid tools or in comfort with
spoken English, assessment changes may create uneven burdens.

Finally, several instructors questioned whether current practices provide valid
evidence of learning. P10, who manages program-level assessment in her
department, noted that colleagues' claims about GenAI ``working'' rested on
perceptions and student enthusiasm rather than on validated measures, and
emphasized the need to assess learning outcomes at the program level. Across
interviews, instructors acted under time pressure and partial observability;
they relied on ambient evidence rather than orchestrated evaluations when
judging whether students both used GenAI as intended and learned the target
skills.

\begin{insights}
  Instructors operated with partial observability, inferring behavior from
  informal, ambient signals. Credit on course assignments tended to shift from
  correctness toward explanation, communication, and judgment. These moves raise
  open questions about validity, reliability, feasibility, and equity, and point
  to a need for ways to capture both how students use GenAI and what they can do
  without it.
\end{insights}

\subsection{Lack of Resources}
\label{sec:resources}

\begin{figure}
  \centering
  \include{lackofresources}
  \caption{\added{A substantial number of our survey respondents expressed that they
      did not have the resources they needed to implement GenAI, which was
      especially prominent for MSI instructors (a.i, a.ii). MSI instructors were
      also more likely to have heavier teaching loads (b.i, b.ii).}
    Instructors generally perceived a lack of sufficient time to learn and
    integrate GenAI (c), and that their institutions did not support this work
    (d). Instructors responded on a 5-point scale. The number of responses for
    each option is displayed within each bar.}

  \Description{Stacked bar plots showing responses to four survey questions:
    (a) Do you have access to sufficient resources to implement GenAI
    effectively?, (b) What is your current teaching load per academic term (on
    average)? (c) Do you feel you have the time to learn and integrate GenAI
    into your teaching, given your current workload?, (d) How well does your
    institution support faculty in adopting GenAI? }
  \label{fig:lackofresources}
\end{figure}

\added{Adapting courses for GenAI required resources that many surveyed
  computing instructors lacked. When asked whether they had sufficient resources
  like funding, training, and tools to implement GenAI effectively, 53\%
  responded No. This pattern was more pronounced among MSI instructors: 62\%
  responded No, compared to 43\% of non-MSI
  instructors~(\fig{fig:lackofresources}a.i, a.ii). Teaching load may help
  explain this difference. MSI instructors were more likely to teach three or
  more courses per term (70\% vs.\ 54\% for
  non-MSIs)~(\fig{fig:lackofresources}b.i, b.ii). In the most extreme cases, 10
  respondents reported teaching six or more courses per academic term; all were
  from MSIs.}

Instructors also expressed a lack of time: when asked if they had time to learn
and integrate GenAI given their current workload, 62\% of surveyed instructors
responded Strongly Disagree, Disagree, or Neither~(\fig{fig:lackofresources}c).
When asked how well their institution supports faculty in adopting emerging
technologies like GenAI, a nearly identical 63\% of faculty responded Very
Poorly, Poorly, or Neither~(\fig{fig:lackofresources}d).

\added{To understand the kinds of resources actually required to adapt a course,
  we examined the resources that our interviewees had when conducting emergency
  pedagogical design.}
We found that our interviewees dedicated substantial amounts of time, staff
hours, and sometimes even their own money in the process of emergency
pedagogical design. For example, bespoke infrastructure required ongoing
development and maintenance. As mentioned in \sec{sec:student-ai}, P04 deployed
a custom chatbot and P09 built an AI-graded explanation workflow; both of these
systems required design, implementation, bug fixes, and upkeep during the term.
These two instructors reported that these systems were only possible with
financial support to hire software engineers and PhD students.

Whole-course redesign also demanded time. Only two interviewees decided to
revise a majority of their course materials~(P05, P08). To accomplish this, both
interviewees adopted course materials that were built by others rather than
developing them from scratch, noting that it would be infeasible to revise an
entire course's worth of curriculum on their own.

Alternative assessments often meant a shift away from autograded work towards
qualitative work, which also required more course staff to manage. As mentioned
in \sec{sec:assessment}, P02 reorganized his course around weekly stand-up
meetings with every student. To handle this for his course of 300 students, he
needed ``a lot of TAs!''~(P02). In practice, this course needed to recruit and
train over 50 course staff members to cover approximately 300 students, a 6:1
staff to student ratio. Others emphasized the hand-grading burden for creative
work, especially in courses that traditionally relied on autograder tests~(P05,
P08, P13).

Direct costs varied. Some instructors reported that their institutions had
agreements with GenAI companies to provide the latest models at no cost to their
students~(P02, P13). Others spent their own money out-of-pocket in order to
support student-AI usage for custom AI tools~(P06, P12). Instructors also
pointed to external funds to offset API or engineering costs (e.g., P04, P08,
P09).

As counterexamples to this pattern, some instructors reached their goals with
off-the-shelf tools and minimal course changes. For example, P01 primarily
guided students to use free off-the-shelf tools to generate example inputs for
algorithms~(P01). Others taught explicit prompting practices without building
bespoke systems (e.g., P03, P07).

What is notable about our interviewees is that they taught at most two courses
per academic term, and many had advantages such as partnerships, external funds,
curriculum materials, or the ability to hire many course staff members. This
pattern is striking: substantial attempts to change student-AI interaction were
made by instructors with lighter teaching loads and access to staffing or
funding.
\added{These accounts highlight the kinds of resources -- time, staffing,
  funding, and infrastructure -- that enabled our interviewees to make
  substantial course changes. Yet our survey data suggests that most computing
  instructors, especially those at MSIs who reported heavier teaching loads and
  less institutional support, do not have comparable resources. Because we view
  our survey sample as a more representative of the constraints faced by
  computing faculty, this gap indicates that many computing instructors may be
  unable to adopt the kinds of course adaptations described by our interviewees
  without additional support.}

\removed{For example, in contrast to our interviewees, a higher
  proportion of our survey respondents had heavier teaching loads: 61\% taught
  three or more courses per academic term. Among instructors at MSIs, this share
  rose to 70\%. In the most extreme cases, 10 respondents reported teaching six
  or more courses per academic term; all were from MSIs. As one might expect, a
  similar proportion expressed a lack of resources: when asked if they had time
  to learn and integrate GenAI given their current workload, 62\% of surveyed
  instructors responded Strongly Disagree, Disagree, or
  Neither~(\fig{fig:lackofresources}a). When asked how well their institution
  supports faculty in adopting emerging technologies like GenAI, a nearly
  identical 63\% of faculty responded Very Poorly, Poorly, or
  Neither~(\fig{fig:lackofresources}b). These results underscore both resource
  requirements and constraints that can prevent instructors from making fruitful
  attempts to improve student-AI interactions.}

\begin{insights}
  \added{Instructors who successfully adapted courses for GenAI tended to have
    advantages like lower teaching load, staffing, and funding. Instructors who
    lacked these resources, especially those at MSIs, found it more difficult to
    effectively engage in emergency pedagogical design.}

  \removed{Across interviews, the binding constraints were clear and separable: time to
    design and maintain changes, staffing to run labor-intensive assessments, and
    money for model access. Instructors who redesigned assessments or built
    tooling tended to have lighter teaching loads and additional support. Survey
    results align with this picture: most faculty reported heavier teaching loads
    (especially at MSIs) and also less capacity to integrate GenAI into their
    course materials.}

\end{insights}

\section{Discussion}
\label{sec:disc}

Our findings position emergency pedagogical design as a distinct design setting
for HCI. Emergency pedagogical design arises when instructors must shape
student-AI interactions under four conditions: reactive timing, indirect control
over commercial tools, partial observability of day-to-day behavior, and time
pressure to act. Without the time needed to conduct large-scale evaluations,
instructors chose to rely on ambient evidence gathered from informal student
observations. We expect that emergency pedagogical design will become less
necessary as GenAI tools stabilize, policies align across courses, and
validated measures of student-AI behavioral alignment are established. In the
meantime, however, there is a rare opportunity for HCI research, academic
institutions, and funding agencies to impact how instructors approach student-AI
interactions in the years to come.

\subsection{Open Questions for HCI Research}
\label{sec:disc-hci}

\added{As outlined by the five barriers of our research
  findings~(\sec{sec:results}), building effective courses is a challenging
  design problem made even more complicated as instructors are now trying to
  influence student-AI interactions. Some barriers, especially Implementation
  Challenges~(\sec{sec:student-ai}) and Assessment
  Misfit~(\sec{sec:assessment}), suggest open questions for future HCI research
  to support instructors in emergency pedagogical design which we pose here for
  the research community.}

\removed{We synthesize our findings into open questions for future HCI research to
  support instructors in emergency pedagogical design.}

\begin{itemize}
  \item Defining traits of effective student-AI interactions: Our interviewees
    expressed a desire to improve how students interact with GenAI tools, and
    often had clear ideas about undesirable behaviors (such as copy-pasting code),
    but struggled to articulate what constitutes more effective or desirable
    student-AI interaction. How can we define a taxonomy of student-AI
    interactions? How can HCI research help instructors translate learning
    outcomes into specific, desirable, and undesirable types of student-AI
    interactions?

  \item Understanding effects of Policy Crosswinds~(\sec{sec:administrative}):
    Students frequently encounter divergent policies regarding GenAI usage
    across different courses. How do students respond to inconsistent policies,
    such as one course requiring AI suggestions to be disabled while another
    encourages their use? What factors influence whether students adhere to
    these varying policies?

  \item Designing desirable course-aligned tools: Since students can easily
    switch to general-purpose chatbots, what affordances and incentives make
    course-specific or course-aligned tools the preferred option for students
    without relying on coercion?

  \item Understanding student-AI behaviors with commercial tools: Instructors
    typically have only partial visibility into how students use commercial
    GenAI tools, relying on proxies to infer behavior, such as scores on a
    proctored quiz. How reliable are current instructor proxies for
    understanding student-AI usage? Can we design more accurate proxies that
    support early detection of undesirable interactions while also preserving
    student privacy? How can these proxies inform instructor decision-making?

  \item Understanding student-AI behaviors with bespoke tools: Instructors who
    develop custom wrappers around GenAI tools theoretically have greater
    insight into student interactions, but often find it difficult to analyze
    large volumes of interaction data. How can tools help instructors make sense
    of student-AI interactions that are directly observable (e.g., through
    concept induction~\cite{lamConcept2024})? How can practices from machine
    learning monitoring be adapted to educational contexts without requiring
    instructors to manage extensive infrastructure or acquire new technical
    expertise?

  \item Supporting adoptability for instructors: Given that instructors face
    significant time, staffing, and financial constraints, how can we design
    tools and interventions that are easy to adopt, require minimal disruption
    to existing course materials, and remain low-cost for instructors?

  \item Making use of ambient evidence: Instructors often rely on informal
    signals such as office hour conversations and staff reports to gain insight
    into student behavior. How can tools help capture a greater quantity and
    quality of ambient data, and support instructors in acting on this
    information? How closely can we approximate the benefits of detailed user
    studies using interactions that already occur, such as office hours and
    online forum posts?

  \item Preserving student privacy: Many approaches to understanding student-AI
    interactions require collecting additional data about student behaviors,
    which risks capturing sensitive information. What types of aggregates and
    visualizations can provide instructors with useful visibility (such as usage
    summaries, failure modes, and guardrail breaches) without exposing sensitive
    content? What consent and governance models are appropriate for educational
    settings?
\end{itemize}

\subsection{Recommendations for Academic Institutions and Funders}
\label{sec:disc-recs}

\begin{figure}
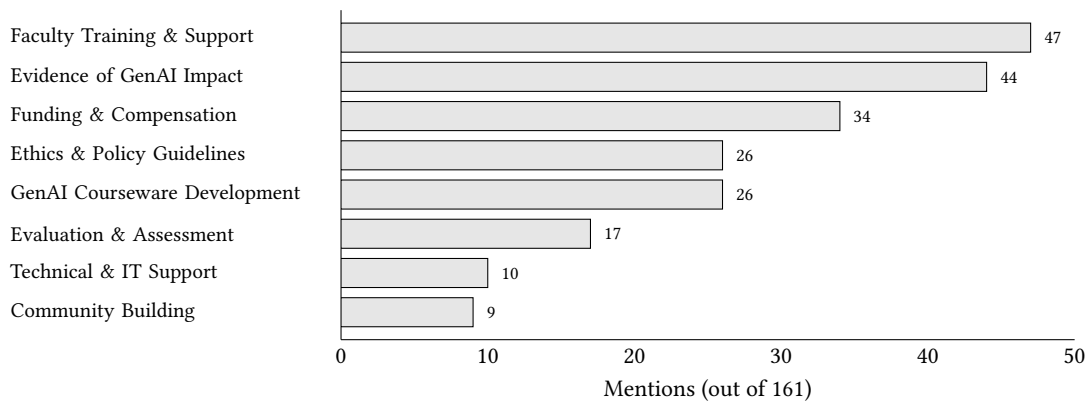

  \centering
  \include{barchart}
  \caption{Responses to the question ``In your opinion, what kinds of support
    would be most helpful to address faculty resistance to teaching GenAI?''
    Survey respondents were allowed to choose up to three options from a
    pre-defined list. Eight most common responses shown, with counts displayed to
    the right of each bar.}
  \Description{Bar chart showing responses to the question ``In your opinion, what kinds of support
    would be most helpful to address faculty resistance to teaching GenAI?''.
    The three most common responses were:
    Faculty Training \& Support (47 mentions);
    Evidence of GenAI Impact (44 mentions); and
    Funding \& Compensation (34 mentions).
  }
  \label{fig:barchart}
\end{figure}

When we began recruiting interviewees for this study, we initially expected that
it would be straightforward to find computing instructors who had updated their
course materials in response to GenAI, given that more than two years had passed
since the release of prominent tools like ChatGPT and GitHub Copilot. However,
we quickly discovered that our assumption was misplaced: it was far easier to
find instructors who had only revised their course policies, rather than those
who had made substantive updates to assignments, assessments, or instructional
content. One possibility is that while the conversation around GenAI in
education is widespread, the actual integration of GenAI into course materials
remains relatively rare.

Our interview findings further revealed that the work of emergency pedagogical
design is both time-consuming and largely unrecognized by institutions. This
burden is exacerbated by the absence of established pedagogical guidelines and
best practices for GenAI instruction, requiring instructors to craft custom
solutions often from scratch. For example, none of our interviewees reported
direct support such as teaching relief, additional course staff, or extra
compensation for the substantial work they performed to redesign materials or
assessments. This lack of institutional recognition and support meant that
instructors shouldered the burden of adaptation on top of their routine teaching
responsibilities, often working in isolation and without formal acknowledgment.
Survey results extended these findings: instructors expressed a desire to
improve student-AI interactions by updating their course materials, yet reported
that they were prevented from doing so by resource constraints, a challenge
especially pronounced at MSIs with heavier teaching loads and more limited
access to staff or funding~(\sec{sec:resources}). These patterns raise equity
concerns, as there is a real risk that well-resourced institutions and
instructors will be able to adapt quickly and effectively to the challenges and
opportunities of GenAI, while less-resourced institutions, particularly those
serving underrepresented student groups, may fall further behind. If only the
most privileged institutions can afford to update their curricula and
assessments, GenAI could reinforce existing disparities in student outcomes and
preparedness.

Not all of the issues surfaced by our study can be addressed solely through HCI
research or technical interventions; \added{barriers like Fragmented
  Buy-in~(\sec{sec:personal}), Policy Crosswinds~(\sec{sec:administrative}), and
  Lack of Resources~(\sec{sec:resources}) are also structural or institutional in
  nature}. In light of these findings, we offer several recommendations for
academic institutions and funders to more effectively support instructors as
they design better student-AI interactions. Academic institutions and
departments should acknowledge that emergency pedagogical design represents
additional work that instructors are taking on while still needing to manage
their routine teaching responsibilities.
We suggest reframing policy changes such as blanket bans on student GenAI usage
as a form of emergency pedagogical design: instructors who impose such bans are
still trying to reduce undesired student-AI interactions via course policy and
face similar challenges in assessing whether these policies have actually
succeeded in influencing student behavior. In this reframing, nearly all
instructors are engaged in emergency pedagogical design of some kind, and we
recommend that academic institutions reward instructors who wish to make
improvements for faculty across their departments and schools. Departments
should make establishing shared policies a priority to reduce the potential
harms of policy crosswinds, especially on students. Institutions should provide
resources for instructors who are leaders within their institutions and
communities, potentially in the form of teaching reliefs, the ability to hire
additional course staff, and stipends for hosting training workshops for other
faculty. Where less resources are available, institutions can recognize
even individual, small-scale interventions and facilitate faculty share-outs to
mitigate fragmented buy-in.

Another notable observation from our interviews was that few interviewees were
aware of how other instructors were making changes to their curriculum and
assessing the effectiveness of these changes outside of their home departments.
To address this, we recommend funding organizations to support efforts to create
communities of practice where instructors can share preliminary work, curriculum
and approaches. This could fulfill a pressing need: when survey respondents
were asked what kinds of support would be most helpful to address faculty
resistance to teaching GenAI, faculty training and support was the most
requested resource, followed closely by evidence of GenAI
impact~(\fig{fig:barchart}).
Although academic research venues remain an important way to share peer-reviewed
experiences, the length of the publication cycle is long, and instructors can
benefit even from informal discussions about work-in-progress approaches. Since
HCI research prioritizes novelty and academic departments prioritize local
impact, we encourage funders to provide incentives for HCI researchers to
produce tools and techniques that can be easily adopted at scale, and for
instructors to broadcast and receive credit for innovative approaches to
emergency pedagogical design that can be more agile than publishing
peer-reviewed papers. All of these recommendations and suggestions reflect a
continued need for administrators, department chairs, and funders to support
faculty as they engage in emergency pedagogical design.

\subsection{Learning from Emergency Remote Teaching}
\label{sec:disc-ert}

\added{Less than a decade ago, instructors also had to make sudden sweeping
  changes to their teaching in response to COVID-19. What are some lessons
  learned from emergency remote teaching that might help address the challenges
  that computing instructors now face when conducting emergency pedagogical
  design? In our view, one of the more helpful perspectives taken during
  COVID-19 was drawing a clear distinction between online learning, where a
  course is specifically designed for an online modality, and emergency remote
  teaching, where a course is offered online ``in a hurry, with bare minimum
  resources, and scant time''~\cite{hodgesdifference2020}. In this paper, we
  argue for a similar distinction. While decades of research have examined how
  to integrate AI into pedagogy (for example, via intelligent tutoring
  systems~\cite{andersonCognitive1995}), this research has also shown that
  effectively using AI often takes forethought and careful design. As was the
  case for emergency remote teaching, we view the lack of resources and time
  that instructors now face in emergency pedagogical
  design~(\sec{sec:resources}) as an indicator that emergency pedagogical design
  should be treated as distinct from designing AI-integrated courses. }

\added{Another outcome of emergency remote teaching was that instructors became
  more familiar with techniques for online learning and blended
  learning~\cite{singhCombining2021} and found them valuable even after a return
  to in-person instruction~\cite{adedoyinCovid192023}. If this pattern also
  holds for emergency pedagogical design, we might expect that Fragmented
  Buy-in~(\sec{sec:personal}) and Policy Crosswinds~(\sec{sec:administrative})
  will become less influential as more instructors will learn about and use
  GenAI tools for instruction. However, this will be made more possible as we
  solve current problems in implementation~(\sec{sec:student-ai}) and
  assessment~(\sec{sec:assessment}) of student-AI interactions, and thus we
  believe that HCI research has a critical role to play in discovering effective
  design practices for the future.}

\section{Conclusion}
\label{sec:conclusion}

We studied how computing instructors adapted courses to widespread GenAI use
through interviews ($n$=13) and a survey ($n$=169), framing this work as
\emph{emergency pedagogical design}: reactive, indirect efforts to shape
student-AI interactions under time pressure and partial observability. We
characterized the properties of emergency pedagogical design and five barriers
that instructors face in this work: fragmented buy-in, policy crosswinds,
implementation challenges, assessment misfit, and lack of resources. We
positioned emergency pedagogical design as a design setting for HCI where
iteration relies on ambient evidence. Looking ahead, we encourage HCI
researchers, tool builders, institutions, and funding agencies to help shift
from emergency responses to routine practice by providing instructors with new
knowledge and tools to improve student-AI interactions, more robust evaluation
methods, time and recognition for instructor efforts, and financial support for
access and maintenance.

\begin{acks}
    We thank Google for their generous support for this project. Large language models were used to revise the text for clarity, spelling, and grammar.
\end{acks}

\bibliography{sam-refs}

\end{document}

%% file: tab-participants.tex
\begin{table*}
  \caption{We interviewed 13 computing instructors to understand their
    experiences conducting emergency pedagogical design for their courses.
    Gender = Male (M), Female (F), Non-binary (NB). Years = number of years of
    experience as a full-time instructor. \added{Each course listed in the
      ``Courses discussed'' column is marked with (L) to denote a lower-division
      course or (U) for an upper-division course.}}
  \label{tab:participants}
  \centering
  \begin{tabularx}{0.85\linewidth}{l l l r l l}
    \toprule
    ID  & Gender & Country     & Years & Type of university     & Courses discussed                           \\
    \midrule
    P01 & F      & US          & 12    & Public PhD-granting    & Algorithms (U)                              \\
    P02 & M      & US          & 26    & Private undergrad-only & Intro to programming (L)                    \\
    P03 & M      & Canada      & 5     & Public PhD-granting    & Data structures (L)                         \\
    P04 & M      & US          & 6     & Public PhD-granting    & Intro to programming (L), Software eng. (U) \\
    P05 & M      & US          & 23    & Public PhD-granting    & Intro to programming (L)                    \\
    P06 & NB     & US          & 11    & Private PhD-granting   & Intro to programming (L)                    \\
    P07 & M      & US          & 28    & Private PhD-granting   & Software design (U)                         \\
    P08 & F      & US          & 9     & Public PhD-granting    & Intro to programming (L)                    \\
    P09 & M      & New Zealand & 12    & Public PhD-granting    & Intro to programming (L)                    \\
    P10 & F      & US          & 22    & Public PhD-granting    & Software engineering (U)                    \\
    P11 & F      & US          & 9     & Public PhD-granting    & Intro to programming (L)                    \\
    P12 & M      & US          & 14    & Public PhD-granting    & Web development (U)                         \\
    P13 & M      & UK          & 42    & Public PhD-granting    & Cybersecurity (L)                           \\
    \bottomrule
  \end{tabularx}
\end{table*}

%% file: tab-results-overview.tex
\begin{table*}
  \caption{\added{Overview of the five barriers that emerged from our analysis
      of interview and survey data. The barriers summarize recurring constraints
      computing instructors encountered while engaging in emergency pedagogical
      design.}
    \removed{Summary of the five barriers in engaging in emergency pedagogical
      design, surfaced through our interviews with computing instructors.}
  }
  \label{tab:results-overview}
  \centering
  \begin{tabularx}{\linewidth}{@{} >{\raggedright}p{0.17\textwidth} X X @{}}
    \toprule
    Barrier
     &
    Description
     &
    Representative Quote
    \\
    \midrule
    Fragmented Buy-In (\sec{sec:personal})
     &
    Instructors have limited personal bandwidth and faced mixed reactions from
    colleagues in their departments.
     &
    ``For years, I've been trying to get my department to at least talk about
    getting students to use GenAI better! And we really haven't, amazingly
    enough.''~(P07)
    \\
    \addlinespace[1em]
    Policy Crosswinds (\sec{sec:administrative})
     &
    Policies around GenAI could be dramatically different for individual courses,
    which led to student confusion.
     &
    ``I have to keep telling my students that using ChatGPT is not cheating in
    my class.  They're scared to use ChatGPT in front of me because other
    classes don't allow it.''~(P01)
    \\
    \addlinespace[1em]
    Implementation Challenges (\sec{sec:student-ai})
     &
    Lacking established pedagogical guidelines, instructors implemented
    experimental, bespoke approaches to shape student-AI interactions.
     &
    ``Honestly, I'm still trying to figure it out, but I'm confident that
    introducing GenAI isn't something we should postpone. The longer we wait,
    the more bad habits students can accrue.''~(P08)
    \\
    \addlinespace[1em]
    Assessment Misfit (\sec{sec:assessment})
     &
    The efficacy of interventions on both student-AI interactions
    and student learning was difficult to measure, even when telemetry data
    was available.
     &
    ``I have hundreds of student logs [with my course's custom chatbot], but I
    don't have the time to analyze them all.''~(P04)
    \\
    \addlinespace[1em]
    Lack Of Resources (\sec{sec:resources})
     &
    Instructors had to work within strict time, staffing, and financial
    constraints which limited their ability to try new approaches.
     &
    ``Without [external] financial support, these changes simply wouldn't have
    happened.''~(P09)
    \\
    \bottomrule
  \end{tabularx}
\end{table*}

%% file: fragmentedbuyin.tex
\begin{tikzpicture}
  \begin{axis}[
      width=0.75\linewidth, height=6cm,
      xbar stacked, xmin=0, xmax=100,
      y=1.2cm, bar width=0.5cm,
      ymin=-0.5, ymax=2.5,
      ytick=data,
      yticklabels={
          {(a) How open are you to adopting GenAI technologies in your curriculum?},
          {(b) How open do you think your colleagues are to adopting GenAI in their curriculum?},
          {(c) My colleague's resistance to GenAI is a barrier to wider adoption within the department.}
        },
      y tick label style={text width=0.33\linewidth, align=left, font=\footnotesize},
      xtick={0,25,50,75,100},
      xticklabel={\pgfmathprintnumber{\tick}\%},
      xticklabel style={font=\footnotesize, yshift=5pt},
      tick style={draw=none}, axis line style={draw=none},
      ytick style={draw=none},
      enlarge y limits=0.02,
      legend style={at={(0.5,-0.38)}, anchor=north, draw=none, fill=gray!10,
          rounded corners, font=\footnotesize, cells={anchor=west}},
    ]
    \addplot+[xbar, fill=likert1,   draw opacity=0] coordinates {(1.9,2)  (4.0 ,1) (15.4,0)};
    \addplot+[xbar, fill=likert2,draw opacity=0] coordinates {   (5.0,2)  (17.2,1) (28.2,0)};
    \addplot+[xbar, fill=likert3,  draw opacity =0] coordinates { (12.5,2) (51.0,1) (33.6,0)};
    \addplot+[xbar, fill=likert4,  draw opacity=0] coordinates { (30.6,2) (20.5,1) (17.4,0)};
    \addplot+[xbar, fill=likert5, draw opacity=0] coordinates {  (50.0,2) (7.3 ,1) (5.4 ,0)};
  \end{axis}
  \node at (.75,3.5) {\footnotesize Not at all open};
  \node at (.75,2.3) {\footnotesize Not at all open};
  \node at (.89,1.1) {\footnotesize Strongly disagree};
  \node at (4.25,3.5) {\footnotesize Very open};
  \node at (4.25,2.3) {\footnotesize Very open};
  \node at (4,1.1) {\footnotesize Strongly agree};
\node at (0.19,3.07) {\footnotesize 8};
\node at (0.598,3.07) {\footnotesize 20};
\node at (1.71,3.07) {\footnotesize 49};
\node at (3.65,3.07) {\color{white} \footnotesize 89};

\node at (0.092,1.87) {\color{white} \footnotesize 6};
\node at (0.552,1.87) {\footnotesize 26};
\node at (2.3,1.87) {\footnotesize 77};
\node at (3.95,1.87) {\footnotesize 31};
\node at (4.57,1.87) {\color{white} \footnotesize 11};

\node at (0.38,.67) {\color{white} \footnotesize 23};
\node at (1.4,.67) {\footnotesize 42};
\node at (2.91,.67) {\footnotesize 50};
\node at (4.13,.67) {\footnotesize 26};
\node at (4.63,.67) {\color{white} \footnotesize 8};
\end{tikzpicture}

%% file: policycrosswind.tex
\begin{tikzpicture}
  \begin{axis}[
      width=0.8\linewidth, height=8cm,
      xbar stacked, xmin=0, xmax=100,
      y=1.2cm, bar width=0.5cm,
      ymin=-0.5, ymax=3.5,
      ytick=data,
      yticklabels={
          {(a) Knowing that students may use GenAI regardless of policies,
              how likely are you to make changes to your curriculum?},
          {(b) How do institutional policies impact your ability to integrate
              GenAI into the curriculum?},
          {\added{(c) Unequal access to paid generative AI tools among students could create or exacerbate
                disparities in learning outcomes \& workforce preparedness.}}
        },
      y tick label style={text width=0.30\linewidth, align=left, font=\footnotesize},
      xtick={0,25,50,75,100},
      xticklabel={\pgfmathprintnumber{\tick}\%},
      xticklabel style={font=\footnotesize, yshift=5pt},
      tick style={draw=none}, axis line style={draw=none},
      ytick style={draw=none},
      enlarge y limits=0.02,
      legend style={at={(0.5,-0.38)}, anchor=north, draw=none, fill=gray!10,
          rounded corners, font=\small, cells={anchor=west}},
    ]
\addplot+[xbar, fill=likert1, draw opacity=0]
  coordinates {(0.6,3)(17.9,1.5)(3.11,0)};

\addplot+[xbar, fill=likert2, draw opacity=0]
  coordinates {(3.2,3)(20.5,1.5)(3.73,0)};

\addplot+[xbar, fill=likert3, draw opacity=0]
  coordinates {(10.8,3)(35.9,1.5)(14.91,0)};

\addplot+[xbar, fill=likert4, draw opacity=0]
  coordinates {(37.6,3)(17.9,1.5)(28.57,0)};

\addplot+[xbar, fill=likert5, draw opacity=0]
  coordinates {(47.8,3)(7.7,1.5)(49.69,0)};

  \end{axis}

  \node at (0.67,4.7) {\footnotesize Very unlikely};
\node at (4.65,4.7) {\footnotesize Very likely};

\node at (0.79,2.9) {\footnotesize Very negatively};
\node at (4.45,2.9) {\footnotesize Very positively};

\node at (0.88,1.1) {\footnotesize Strongly disagree};
\node at (4.45,1.1) {\footnotesize Strongly agree};

\node at (0.1,4.28) {\footnotesize 5};
\node at (0.46,4.28) {\footnotesize 17};
\node at (1.75,4.28) {\footnotesize 59};
\node at (4,4.28) {\color{white} \footnotesize 75};

\node at (0.45,2.48) {\color{white} \footnotesize 28};
\node at (1.45,2.48) {\footnotesize 32};
\node at (2.9,2.48) {\footnotesize 56};
\node at (4.4,2.48) {\footnotesize 28};
\node at (4.98,2.48) {\color{white} \footnotesize 12};

\node at (0.0736,0.68) {\color{white} \footnotesize 5};
\node at (0.25,0.68) {\footnotesize 6};
\node at (0.71,0.68) {\footnotesize 24};
\node at (1.9,0.68) {\footnotesize 46};
\node at (3.97,0.68) {\color{white} \footnotesize 80};
\end{tikzpicture}

%% file: implementationchallenges.tex
\begin{tikzpicture}
  \begin{axis}[
      width=0.78\linewidth, height=6cm,
      xbar stacked, xmin=0, xmax=100,
      y=1.2cm, bar width=0.5cm,
      ymin=-0.5, ymax=2.5,
      ytick=data,
      yticklabels={
          {(a) How important do you think it is to integrate GenAI tools into your curriculum?},
          {(b) How frequently do you use GenAI tools in course activities?}
        },
      y tick label style={text width=0.33\linewidth, align=left, font=\footnotesize},
      xtick={0,25,50,75,100},
      xticklabel={\pgfmathprintnumber{\tick}\%},
      xticklabel style={font=\footnotesize, yshift=5pt},
      tick style={draw=none}, axis line style={draw=none},
      ytick style={draw=none},
      enlarge y limits=0.02,
      legend style={at={(0.5,-0.35)}, anchor=north, draw=none, fill=gray!10,
          rounded corners, font=\footnotesize, cells={anchor=west}},
    ]
    \addplot+[xbar, fill=likert1, draw opacity=0] coordinates {(1.9,1) (17.1,0)};
    \addplot+[xbar, fill=likert2, draw opacity=0] coordinates {(3.1,1) (20.3,0)};
    \addplot+[xbar, fill=likert3, draw opacity=0] coordinates {(15.5,1)(25.3,0)};
    \addplot+[xbar, fill=likert4, draw opacity=0] coordinates {(32.3,1)(18.4,0)};
    \addplot+[xbar, fill=likert5, draw opacity=0] coordinates {(47.2,1)(19.0,0)};

  \end{axis}
  \node at (1,2.3) {\footnotesize Not at all important};
  \node at (.33,1.1) {\footnotesize Never};
  \node at (4.23,2.3) {\footnotesize Very important};
  \node at (4.47,1.1) {\footnotesize Very often};
  \node at (.17,1.87) {\footnotesize 5};
  \node at (.63,1.87) {\footnotesize 25};
  \node at (1.88,1.87) {\footnotesize 52};
  \node at (3.88,1.87) {\color{white} \footnotesize 76};

  \node at (0.44,.67) {\color{white} \footnotesize 27};
  \node at (1.37,.67) {\footnotesize 32};
  \node at (2.53,.67) {\footnotesize 40};
  \node at (3.65,.67) {\footnotesize 29};
  \node at (4.57,.67) {\color{white} \footnotesize 30};
\end{tikzpicture}

%% file: lackofresources.tex
\begin{tikzpicture}
  \begin{axis}[
      width=0.75\linewidth, height=9.5cm,
      xbar stacked, xmin=0, xmax=100,
      y=1.2cm, bar width=0.5cm,
      ymin=-0.5, ymax=5.5,
      ytick=data,
      yticklabels={
          {\added{(a.i) Do you have access to sufficient resources (funding, training, tools) to implement generative AI effectively?}},
          {\added{(a.ii) Do you have access to sufficient resources (funding, training, tools) to implement generative AI effectively?}},
          {\added{(b.i) What is your current teaching load per academic term (on average)? (MSI)}},
          {\added{(b.ii) What is your current teaching load per academic term (on average)? (non-MSI)}},
          {(c) Do you feel you have the time to learn and integrate GenAI into your teaching, given your current workload?},
          {(d) How well does your institution support faculty in adopting GenAI?}
        },
      y tick label style={text width=0.33\linewidth, align=left, font=\footnotesize},
      xtick={0,25,50,75,100},
      xticklabel={\pgfmathprintnumber{\tick}\%},
      xticklabel style={font=\footnotesize, yshift=5pt},
      tick style={draw=none}, axis line style={draw=none},
      ytick style={draw=none},
      enlarge y limits=0.02,
      legend style={at={(0.5,-0.35)}, anchor=north, draw=none, fill=gray!10,
          rounded corners, font=\footnotesize, cells={anchor=west}},
    ]
    \addplot+[xbar, fill=likert1, draw opacity=0]
  coordinates {(61.6,5.5) (43.1,4.2) (0,3.3) (0,2.2) (9.4,1) (8.1,0)};

\addplot+[xbar, fill=likert2, draw opacity=0]
  coordinates {(0,5.5) (0,4.2) (0,3.3) (0,2.2) (21.4,1) (16.9,0)};

\addplot+[xbar, fill=likert3, draw opacity=0]
  coordinates {(12.8,5.5) (36.9,4.2) (0,3.3) (0,2.2) (31.4,1) (38.1,0)};

\addplot+[xbar, fill=likert4, draw opacity=0]
  coordinates {(0,5.5) (0,4.2) (0,3.3) (0,2.2) (22.0,1) (22.5,0)};

\addplot+[xbar, fill=likert5, draw opacity=0]
  coordinates {(25.6,5.5) (20,4.2) (0,3.3) (0,2.2) (15.7,1) (14.4,0)};

\addplot+[xbar, fill=seq1, draw opacity=0]
  coordinates {(0,5.5) (0,4.2) (31,3.3) (46.3,2.2) (0,1) (0,0)};

\addplot+[xbar, fill=seq2, draw opacity=0]
  coordinates {(0,5.5) (0,4.2) (45.2,3.3) (41.8,2.2) (0,1) (0,0)};

\addplot+[xbar, fill=seq3, draw opacity=0]
  coordinates {(0,5.5) (0,4.2) (11.9,3.3) (11.9,2.2) (0,1) (0,0)};

\addplot+[xbar, fill=seq4, draw opacity=0]
  coordinates {(0,5.5) (0,4.2) (11.9,3.3) (0,2.2) (0,1) (0,0)};

  \end{axis}
  \node at (0.15,7.59) {\footnotesize No};
  \node at (3.25,7.59) {\footnotesize Not sure};
  \node at (4.6,7.59) {\footnotesize Yes};
  \node at (1.6,7.27) {\color{white} \footnotesize 53};
  \node at (3.25,7.27) {\footnotesize 11};
  \node at (4.15,7.27) {\color{white} \footnotesize 22};

  \node at (0.15,6.13) {\footnotesize No};
  \node at (2.93,6.13) {\footnotesize Not sure};
  \node at (4.6,6.13) {\footnotesize Yes};
  \node at (1.09,5.79) {\color{white} \footnotesize 28};
  \node at (2.98,5.79) {\footnotesize 24};
  \node at (4.29,5.79) {\color{white} \footnotesize 13};

  \node at (.55,5.07) {\footnotesize 1-2 courses};
  \node at (2.6,5.07) {\footnotesize 3-4 courses};
  \node at (3.92 ,5.07) {\footnotesize 5-6};
  \node at (4.5 ,5.07) {\footnotesize 6+};
  \node at (.77,4.69) {\footnotesize 26};
  \node at (2.68,4.69) {\footnotesize 38};
  \node at (3.93 ,4.69) {\footnotesize 10};
  \node at (4.49 ,4.69) {\color{white} \footnotesize 10};

  \node at (.55,3.75) {\footnotesize 1-2 courses};
  \node at (3.27,3.75) {\footnotesize 3-4 courses};
  \node at (4.52 ,3.75) {\footnotesize 5-6};
  \node at (1.15,3.37) {\footnotesize 31};
  \node at (3.30,3.37) {\footnotesize 28};
  \node at (4.48 ,3.37) {\color{white}\footnotesize 8};

  \node at (.85,2.32) {\footnotesize Strongly disagree};
  \node at (9.,2.32) {\footnotesize Strongly agree};
  \node at (0.22,1.92) {\color{white} \footnotesize 15};
  \node at (.92,1.92) {\footnotesize 34};
  \node at (2.24,1.92) {\footnotesize 50};
  \node at (3.54,1.92) {\footnotesize 35};
  \node at (4.44,1.92) {\color{white} \footnotesize 25};

  \node at (.55,1.12) {\footnotesize Very poorly};
  \node at (9.3,1.12) {\footnotesize Excellent};
  \node at (0.18,.72) {\color{white} \footnotesize 13};
  \node at (.8,.72) {\footnotesize 27};
  \node at (2.1,.72) {\footnotesize 61};
  \node at (3.57,.72) {\footnotesize 36};
  \node at (4.43,.72) {\color{white} \footnotesize 23};
\end{tikzpicture}

%% file: barchart.tex
\begin{tikzpicture}
  \begin{axis}[
      width=.6\linewidth,
      xbar,
      xmin=0, xmax=50,
      bar width=1.0\baselineskip,
      y=1.75\baselineskip,
      ytick=data,
      symbolic y coords={
        Community Building,
        Technical \& IT Support,
        Evaluation \& Assessment,
        GenAI Courseware Development,
        Ethics \& Policy Guidelines,
        Funding \& Compensation,
        Evidence of GenAI Impact,
        Faculty Training \& Support
      },
      yticklabel style={
        text width=0.47\linewidth,
        align=left,
        anchor=east,
        inner sep=0pt,
        font=\small
      },
      axis x line*=bottom, axis y line*=left,
      tick style={draw=none},
      xtick distance=10, xlabel={Mentions (out of 161)},
      nodes near coords,
      point meta=explicit symbolic,
      every node near coord/.append style={
        font=\footnotesize,
        black,
        anchor=west,
        xshift=2pt
      },
    ]

    \addplot+[draw=black, fill=gray!20] coordinates
      {(9,Community Building)                  [9]
       (10,Technical \& IT Support)            [10]
       (17,Evaluation \& Assessment)           [17]
       (26,GenAI Courseware Development)       [26]
       (26,Ethics \& Policy Guidelines)        [26]
       (34,Funding \& Compensation)            [34]
       (44,Evidence of GenAI Impact)           [44]
       (47,Faculty Training \& Support)        [47]};
  \end{axis}
\end{tikzpicture}